\documentclass[aps,prl,twocolumn,notitlepage,superscriptaddress]{revtex4-1}
\usepackage[colorlinks=true, citecolor=magenta, linkcolor=blue, urlcolor=blue]{hyperref}
\usepackage{graphicx}
\usepackage{amsmath}
\usepackage{amssymb}
\usepackage{amsfonts}
\usepackage{hyperref}
\usepackage{mathtools}
\usepackage{xcolor}
\usepackage{tikz}
\usetikzlibrary{shapes.geometric, arrows}
\usetikzlibrary{decorations.markings}
\usetikzlibrary{decorations.pathmorphing}
\usetikzlibrary{decorations.pathreplacing}

\DeclareMathOperator{\re}{Re}
\DeclareMathOperator{\im}{Im}

\newcommand{\bk}{{\bf k}}
\newcommand{\bA}{{\bf A}}

\allowdisplaybreaks

\begin{document} 

\title{Direct optical probe of magnon topology in two-dimensional quantum magnets}

\author{Emil Vi\~nas Bostr\"om}
\email{emil.bostrom@mpsd.mpg.de}
\affiliation{Max Planck Institute for the Structure and Dynamics of Matter, Center for Free Electron Laser Science (CFEL), Luruper Chaussee 149, 22761 Hamburg, Germany}
\author{Tahereh Sadat Parvini}
\affiliation{Institute of Physics, University of Greifswald, Felix-Hausdorff-Str. 6, Greifswald, 17489, Germany}
\author{James W. McIver}
\affiliation{Max Planck Institute for the Structure and Dynamics of Matter, Center for Free Electron Laser Science (CFEL), Luruper Chaussee 149, 22761 Hamburg, Germany}
\affiliation{Department of Physics, Columbia University, New York, NY 10027}
\author{Angel Rubio}
\affiliation{Max Planck Institute for the Structure and Dynamics of Matter, Center for Free Electron Laser Science (CFEL), Luruper Chaussee 149, 22761 Hamburg, Germany}
\affiliation{Center for Computational Quantum Physics, The Flatiron Institute, 162 Fifth Avenue, New York, NY 10010, United States of America}
\author{Silvia {Viola Kusminskiy}}
\affiliation{Institute for Theoretical Solid State Physics, RWTH Aachen University, 52074 Aachen, Germany}
\affiliation{Max Planck Institute for the Science of Light, Staudtstrasse 2, PLZ 91058 Erlangen, Germany}
\author{Michael A.~Sentef}
\email{michael.sentef@mpsd.mpg.de}
\affiliation{Max Planck Institute for the Structure and Dynamics of Matter, Center for Free Electron Laser Science (CFEL), Luruper Chaussee 149, 22761 Hamburg, Germany}

\date{\today}

\begin{abstract}
Controlling edge states of topological magnon insulators is a promising route to stable spintronics devices. However, to experimentally ascertain the topology of magnon bands is a challenging task. Here we derive a fundamental relation between the light-matter coupling and the quantum geometry of magnon states. This allows to establish the two-magnon Raman circular dichroism as an optical probe of magnon topology in honeycomb magnets, in particular of the Chern number and the topological gap. Our results pave the way for interfacing light and topological magnons in functional quantum devices.
\end{abstract}

\maketitle

%%%%%%%%%%%%%%%%%%%%%%%%%%%%%%%%%%%%%%%%%%%%%%%%%%%%%%%
%%%%%%%%%%%%%%%%%%%%%%%%%%%%%%%%%%%%%%%%%%%%%%%%%%%%%%%
%%%%%%%%%%%%%%%%%%%%%%%%%%%%%%%%%%%%%%%%%%%%%%%%%%%%%%%

%\section*{Introduction}
The study of topological states of matter has recently been extended to systems with bosonic quasi-particles such as magnons, photons, excitons and polaritons~\cite{kondo2019three, Lu2014,Ozawa2019,Karzig2015, wang2018topological,Klembt2018, ozawa2019probing, chaudhary2022shift,Nakata17a,Bostrom20}. The large interest in topological states stems from the hope of utilizing their properties to realize fault-tolerant quantum information devices for large-scale quantum computing. Several predictions exist of how to realize paradigmatic models of topological matter such as the quantum Hall and quantum anomalous Hall effects with magnons~\cite{Nakata17a,Nakata17b,Diaz2019,Bostrom20}. In electronic systems, where the edge state population can be controlled by shifting the chemical potential, e.g., through electrostatic gating, it is straightforward to experimentally verify the topological nature of a given state via conductivity measurements. In contrast, magnon systems lack a chemical potential, and the ground state is usually a Bose-Einstein distribution centered around zero momentum. In order to harness magnon edge states for the realization of stable spintronics devices, it is therefore necessary to find other means of probing the topology of the magnon bands.

The topology of two-dimensional band structures is quantified by their Chern numbers, which are given by an integral of the Berry curvature over the Brillouin zone~\cite{Berry1984}. The Berry curvature can in turn be viewed as the imaginary part of the more general quantum geometric tensor~\cite{Berry1989}, which endows the Hilbert space of quantum states with a Riemannian structure~\cite{Provost1980}. Both the Berry curvature and the quantum metric, the real part of the quantum geometric tensor, are crucial for the understanding of a plethora of physical effects, such as flat band superfluidity~\cite{Peotta2015}, superconductivity~\cite{Julku2016}, orbital magnetic susceptibility~\cite{Gao2014,Rhim2020} and the non-adiabatic anomalous Hall effect~\cite{Bleu2018}. Recently, a connection was found between the quantum geometric tensor and the light-matter coupling in non-interacting fermionic systems~\cite{Topp2021}, as well as between the Berry curvature and angle-resolved photo-emission spectra~\cite{Muechler2016,Schuler2020,Beaulieu2020, yale2016optical}. Although a similar connection for  bosonic systems would allow to optically address the topology of magnon bands, the generalization of these results to boson systems is non-trivial due to the different exchange statistics and transformation properties of the boson operators.

Here, we specifically show that the magnon topology of canted honeycomb antiferromagnets can be probed at zero temperature by the two-magnon Raman circular dichroism (RCD). We demonstrate that the frequency-integrated RCD is tied to the Chern number of the magnon bands, while frequency-resolved measurements of the RCD give access to the size of the topological magnon gap. More generally, we show that the connection between the RCD and the magnon topology follows from a fundamental relation between the light-matter coupling and the quantum geometric tensor for non-interacting boson systems, which we derive. Our results are relevant for a large class of van der Waals (vdW) honeycomb magnets, where the magnon band structure is topological~\cite{Chen2018,Chen2021}.

%%%%%%%%%%%%%%%%%%%%%%%%%%%%%%%%%%%%%%%%%%%%%%%%%%%%%%%
%%%%%%%%%%%%%%%%%%%%%%%%%%%%%%%%%%%%%%%%%%%%%%%%%%%%%%%
%%%%%%%%%%%%%%%%%%%%%%%%%%%%%%%%%%%%%%%%%%%%%%%%%%%%%%%

\begin{figure}
 \includegraphics[width=\columnwidth]{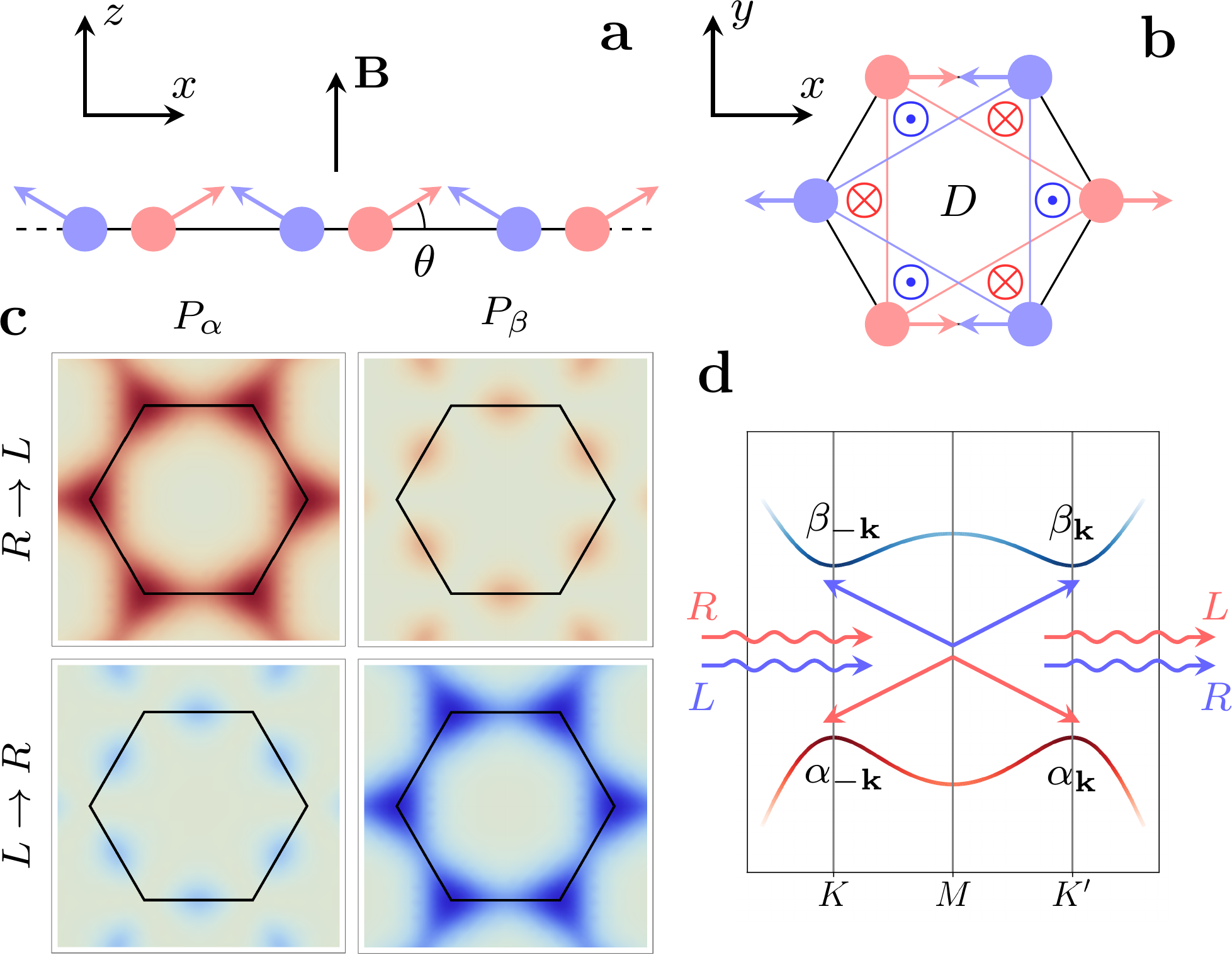}
 \caption{{\bf Magnon band structure and two-magnon Raman scattering in a canted honeycomb antiferromagnet.} {\bf a, b.} Illustration of a canted antiferromagnet on the honeycomb lattice. Panel {\bf a} shows a side view of the system indicating the in-plane N\'eel order, out-of-plane ferromagnetic order and canting angle $\theta$. Panel {\bf b} shows a top view illustrating the direction of the out-of-plane Dzyaloshinskii-Moriya interaction on each next-nearest neighbor bond. {\bf c.} Probabilities $P_{\alpha/\beta}$ for magnon pair creation in the lower/upper magnon branch, via Raman scattering of right- to left-hand polarized light, and vice versa. {\bf d.} Schematic of the two-magnon Raman processes leading to a non-zero circular dichroism: Incident photons of right- or left-handed polarization (red/blue wiggly lines) are scattered into left- or right-handed photons, respectively, while simultaneously creating a magnon pair at the $K$ or $K'$ points (solid lines). For right- to left-handed (left- to right-handed) scattering magnons are predominantely created in the lower (upper) band.}
 \label{fig:equilibrium}
\end{figure}

%\section*{Results}
%\subsection*{Equilibrium canted antiferromagnet}\label{sec:hamiltonian}
The low-energy magnetic properties of monolayer vdW transition-metal (phosphorous) trichalcogenides, such as CrI$_3$, CrCl$_3$, MnPS$_3$ and MnPSe$_3$, are described by a short-range spin Hamiltonian on the honeycomb lattice~\cite{BedoyaPinto2021, sivadas2015magnetic}. To lowest order, this Hamiltonian reads
\begin{align}\label{eq:spin_ham}
 H_0 &= J_{xy} \sum_{\langle ij\rangle} (S_i^x S_j^x + S_i^y S_j^y) + J_z \sum_{\langle ij\rangle} S_i^z S_j^z \\
 &+ D \sum_{\langle\langle ij\rangle\rangle} \nu_{ij} \hat{\bf z} \cdot ({\bf S}_i \times {\bf S}_j) - {\bf B} \cdot \sum_i {\bf S}_i, \nonumber
\end{align}
where $J_{xy}$ and $J_z$ are the in-plane and perpendicular nearest-neighbor exchange interactions, $D$ is the strength of the next-nearest neighbor Dzyaloshinskii-Moriya interaction (DMI), and ${\bf B}$ is an external magnetic field. The coefficients $\nu_{ij}$ arise from electronic virtual hopping processes along isosceles triangles, and take values $\nu_{ij} = \pm 1$ as illustrated in Fig.~\ref{fig:equilibrium}. Although the intrinsic strength of the DMI might be small, a synthetic scalar spin chirality interaction can be induced via circularly polarized lasers~\cite{Claassen2017,Kitamura2017,Bostrom20}, and the DMI can be enhanced through the application of out-of-plane electric fields~\cite{Koyama2018,Ba2021}.

\begin{figure}
 \includegraphics[width=\columnwidth]{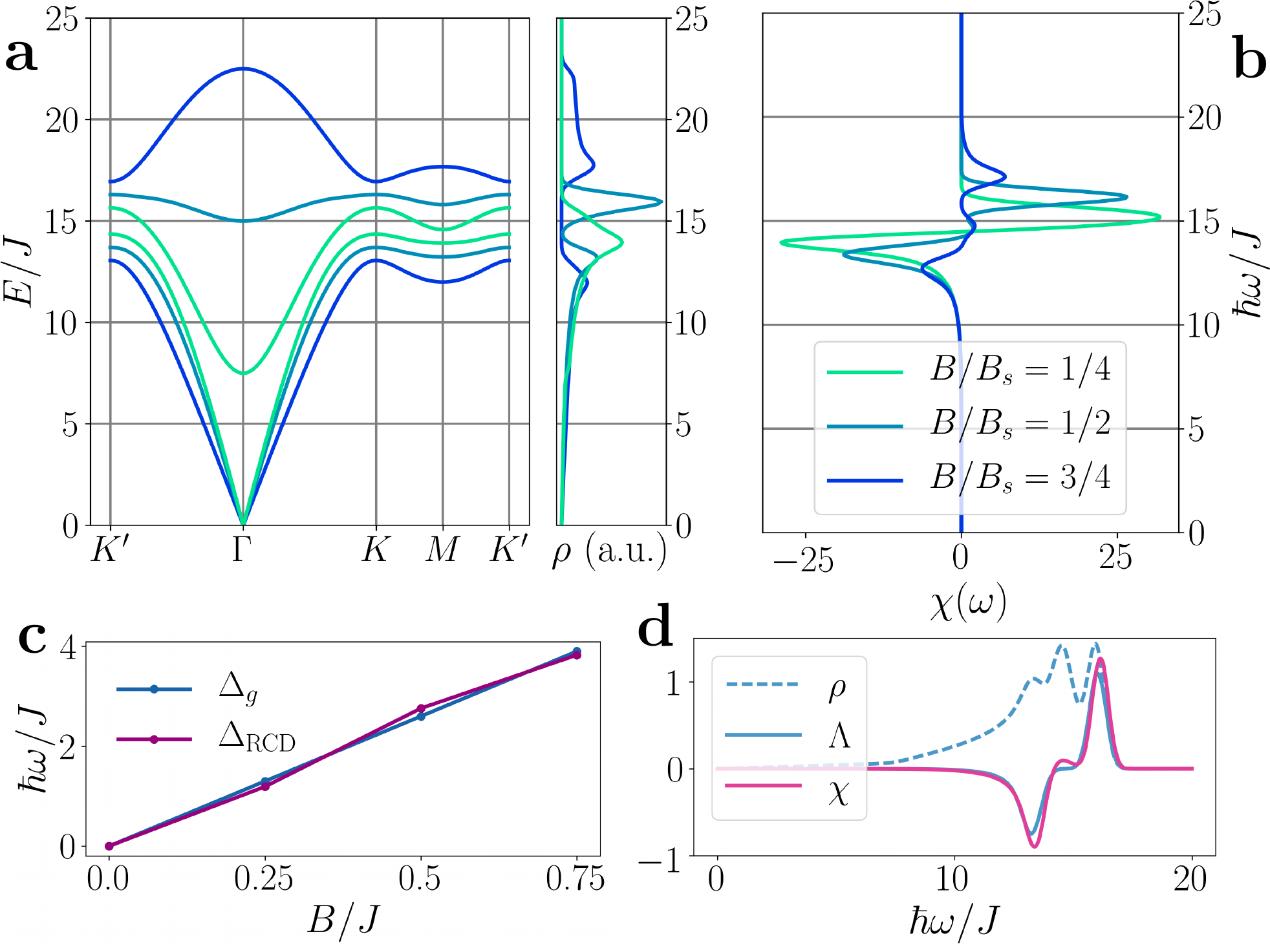}
 \caption{{\bf Raman circular dichroism, topological gap and magnon Berry curvature.} {\bf a.} Magnon band structure and density of states as a function of magnetic field (see {\bf b} for labels). {\bf b.} Frequency-resolved Raman circular dichroism $\chi$ as a function of $\hbar\omega = \hbar\omega_{\rm in} - \hbar\omega_{\rm sc}$. {\bf c.} Magnon gap $\Delta_g$ at $K$ and peak distance $\Delta_{\rm RCD}$ of the Raman circular dichroism as a function of magnetic field. {\bf d.} Two-magnon density of states $\rho$, density of states $\Lambda$ weighted by the magnon Berry curvature, and Raman circular dichroism $\chi$ as a function of photon energy for $B/B_s = 0.5$. In all panels the parameters are $J = 1$, $D/J = 0.1$, $B_s = 6JS$ and $S = 5/2$.}
 \label{fig:frequency}
\end{figure}

For $J_{xy} > J_z$ and an out-of-plane magnetic field ${\bf B} = B \hat{\bf z}$, the ground state of Eq.~\ref{eq:spin_ham} has a canted antiferromagnetic order (see Fig.~\ref{fig:equilibrium}). The Ne\'el vector of the field-free system is taken to lie along the $x$-axis, and will tilt into the $xz$-plane as $B$ increases \cite{Maksimov16, huang2022topological}. Employing a sublattice-dependent Holstein-Primakoff transformation around the local spin axes, the Hamiltonian is given to leading order in $S^{-1}$ by $H_0 = \sum_\bk \Phi_\bk^\dagger \mathcal{H}_{0\bk} \Phi_\bk$ in the basis $\Phi_\bk^\dagger = (a_\bk^\dagger, b_\bk^\dagger, a_{-\bk}, b_{-\bk})$. To diagonalize the Hamiltonian, we perform the Bogoliubov transformation $\Phi_\bk=U_\bk \Psi_\bk$, where $U_\bk$ is a paraunitary matrix, and the resulting magnon energies are denoted $\epsilon_{\bk m}$ ($m = \pm$). The magnon band structure as a function of $B$ interpolates between that of a collinear AFM ($B = 0$) and that of a collinear FM obtained above the saturation field $B_s = 6JS$ (see Fig.~\ref{fig:frequency}).

\begin{figure}
 \includegraphics[width=\columnwidth]{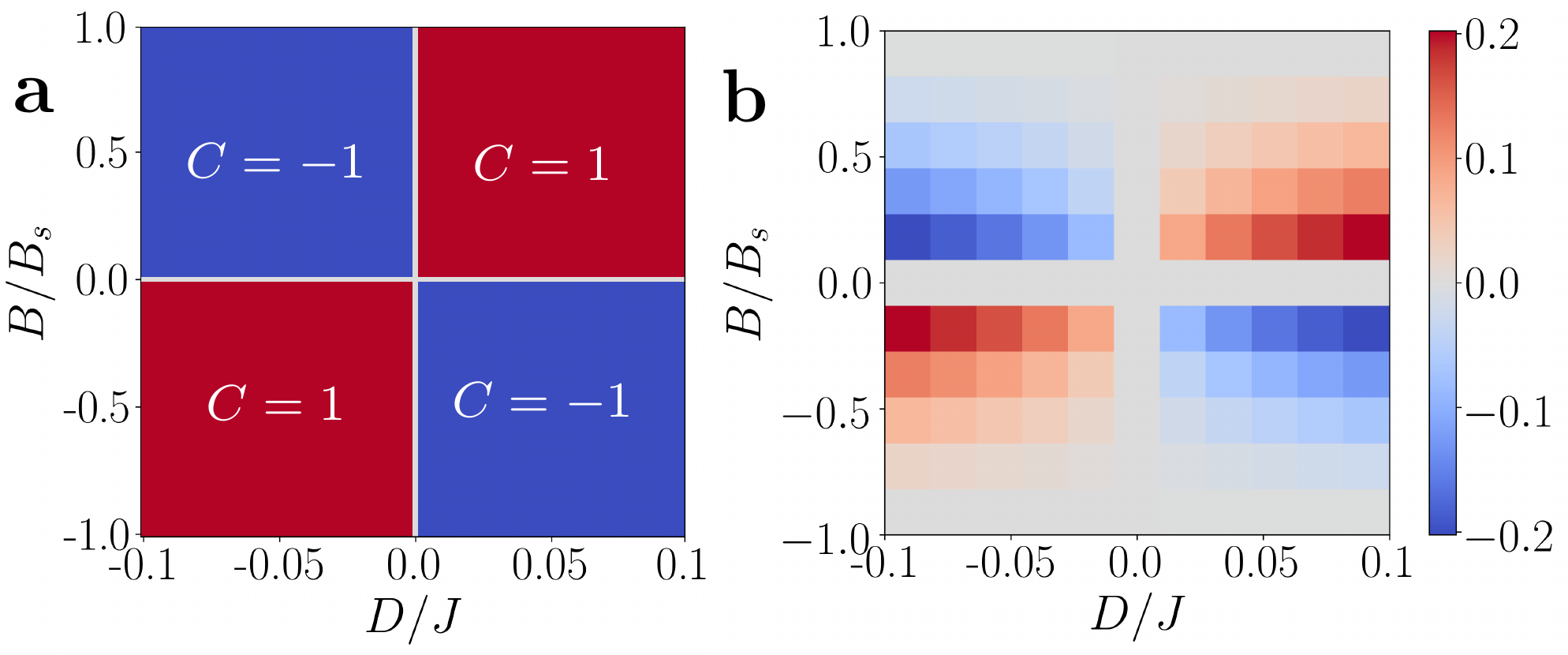}
 \caption{{\bf Raman circular dichroism as a probe of magnon band topology.} {\bf a.} Chern number of the lower magnon band as a function of Dzyaloshinskii-Moriya interaction $D$ and external magnetic field $B$. {\bf b.} Frequency integrated Raman circular dichroism $\chi$. The parameters are $J = 1$, $B_s = 6JS$ and $S = 5/2$.}
 \label{fig:canted}
\end{figure}

The Berry curvature of an antiferromagnet can be written as $\Omega_m(\bk) = \sum_n \Omega_m^{(n)}(\bk)$, where $\Omega_m^{(n)}$ is the contribution of band $n$ to the Berry curvature of band $m$ and is defined by~\cite{Owerre17}
\begin{align}
 \Omega_m^{(n)}(\bk) &= -2\im \frac{[\bar{u}_{m\bk} \tau_z \partial_y \mathcal{H}_{0\bk} u_{n\bk}] [\bar{u}_{n\bk} \tau_z \partial_x \mathcal{H}_{0\bk} u_{m\bk}]}{(\epsilon_{\bk n} - \epsilon_{\bk m})^2}.
\end{align}
Here $u_{m\bk}$ $(\bar{u}_{n\bk}$) is the $m$th column ($n$th row) of the transformation matrix $U_\bk$ ($\bar{U}_\bk = \tau_z U_\bk \tau_z$), and the energies in the denominator are the eigenvalues of the matrix $\tau_z H_0$. The Chern number of band $m$ is given by the Brillouin zone integral $C_m = (2\pi)^{-1} \int_{BZ} d\bk\, \Omega_m(\bk)$. Except for the lines $D = 0$ and $B = 0$, the magnon bands have non-zero Chern numbers and are topological (see Fig.~\ref{fig:canted}). The dominant contribution to the Berry curvature of the magnon bands comes from the $K$ and $K'$ points, where $\Omega_+ = -\Omega_-$.

%%%%%%%%%%%%%%%%%%%%%%%%%%%%%%%%%%%%%%%%%%%%%%%%%%%%%%%
%%%%%%%%%%%%%%%%%%%%%%%%%%%%%%%%%%%%%%%%%%%%%%%%%%%%%%%
%%%%%%%%%%%%%%%%%%%%%%%%%%%%%%%%%%%%%%%%%%%%%%%%%%%%%%%

\begin{table}[t]
    \centering\footnotesize
    \begin{tabular}{|c|c|} \hline
     Mechanism & Physical process \\ \hline\hline
     Aharonov-Casher & Phase accumulation of magnetic moment \\
     Effect & in an electric field \\ \hline
     Peierls & Electric field modulation of virtual electronic \\
     phases & hopping processes \\ \hline
     Inverse Faraday & An effective magnetic field generation by \\
     Effect & the optical spin density \\ \hline
    \end{tabular}
    \caption{{\bf Mechanisms of light-matter coupling.} Summary of the most common light-matter coupling mechanisms and their underlying physical processes.}
    \label{tab:lmc}
\end{table}

\begin{table*}[t]
    \centering\footnotesize
    \begin{tabular}{|p{0.1\textwidth}|p{0.45\textwidth}|p{0.45\textwidth}|} \hline
     & \centering Ferromagnetic & \hspace*{2.25cm} Antiferromagnetic \\ \hline\hline
    \centering $L_{\mu,nn}^{(1)}$ & \centering $- \partial_\mu \epsilon_{n\bk}$ & \hspace*{3.2cm} $- \partial_\mu \epsilon_{n\bk}$ \\ \hline
    \centering $L_{\mu,nm}^{(1)}$ & \centering $- (\epsilon_{m\bk} - \epsilon_{n\bk}) \bar{u}_{n\bk} \partial_\mu u_{m\bk}$ & \hspace*{2.1cm} $- (\epsilon_{m\bk} + \epsilon_{n\bk}) \bar{u}_{n\bk} \partial_\mu u_{m\bk}$ \\ \hline
    \centering $L_{\mu\nu,nn}^{(2)}$ & \centering $\partial_\mu \partial_\nu \epsilon_{n\bk} + 2\sum_l (\epsilon_{n\bk} - \epsilon_{l\bk}) \re\big[ (\bar{u}_{n\bk} \partial_\mu u_{l\bk}) (\bar{u}_{l\bk} \partial_\nu u_{n\bk}) \big]$ & \hspace*{0.3cm} $\partial_\mu \partial_\nu \epsilon_{n\bk} - 2\sum_l (\epsilon_{n\bk} + \epsilon_{l\bk}) \re\big[ ( \bar{u}_{n\bk} \partial_\mu u_{l\bk}) (\bar{u}_{l\bk} \partial_\nu u_{n\bk}) \big]$ \\ \hline
    & \centering $\big[ (\partial_\mu \epsilon_{m\bk} - \partial_\mu \epsilon_{n\bk}) \bar{u}_{n\bk} \partial_\nu u_{m\bk}$ & \hspace*{2.1cm}$\big[ (\partial_\mu \epsilon_{m\bk} + \partial_\mu \epsilon_{n\bk}) \bar{u}_{n\bk} \partial_\nu u_{m\bk}$ \\
    \centering $L_{\mu\nu,nm}^{(2)}$ & \centering $+ \frac{1}{2} (\epsilon_{m\bk} - \epsilon_{n\bk}) \bar{u}_{n\bk} \partial_\mu \partial_\nu u_{m\bk}$ & \hspace*{2.02cm} $+ \frac{1}{2} (\epsilon_{m\bk} + \epsilon_{n\bk}) \bar{u}_{n\bk} \partial_\mu \partial_\nu u_{m\bk}$ \\
    & \centering $- \sum_l (\epsilon_{n\bk} - \epsilon_{l\bk}) (\bar{u}_{n\bk} \partial_\mu u_{l\bk}) (\bar{u}_{l\bk} \partial_\nu u_{m\bk})\big] + (\mu \leftrightarrow \nu)$ & \hspace*{0.68cm} $- \sum_l (\epsilon_{n\bk} + \epsilon_{l\bk}) (\bar{u}_{n\bk} \partial_\mu u_{l\bk}) (\bar{u}_{l\bk} \partial_\nu u_{m\bk})\big] + (\mu \leftrightarrow \nu)$ \\ \hline
    \end{tabular}
    \caption{{\bf Light-matter couplings and quantum geometry.} Linear and quadratic light-matter couplings of interband and pair creation/annihilation processes in terms of the single-magnon energies $\epsilon_{n\bk}$ and Bogoliubov transformation matrices $U_\bk$ with columns $u_{n\bk}$. The vectors $\bar{u}_{n\bk}$ denote the rows of the inverse matrix $U_\bk^{-1}$. In the summations over $l$, care must be taken to assign the correct sign to the energies $\epsilon_{l\bk}$ in accord with their sign as eigenvalues of $\tau_z\mathcal{H}_\bk$.}
    \label{tab:quantum_geometry}
\end{table*}

%\subsection*{Light-matter coupling and quantum geometry}\label{sec:quant_geom}
In presence of an external electromagnetic field, the Hamiltonian acquires a dependence on the vector potential ${\bf A}$. In magnetic systems this dependence can arise from a variety of optomagnetic interactions, the most common of which are the Peierls coupling \cite{shastry1990theory, fleury1968scattering, Bostrom2021}, the Aharonov-Casher effect (ACE) \cite{aharonov1984topological, nakata2014josephson}, and the inverse Faraday effect (IFE) \cite{Ziel1965,Pershan1966}. The microscopic processes underlying these interactions are briefly summarized in Tab.~\ref{tab:lmc}. At optical frequencies the dominant mechanism is two-magnon Raman scattering, where magnon pairs are created at finite $\bk$ with equal and opposite momenta \cite{fleury1968scattering}. In honeycomb antiferromagnets the Raman scattering probability is dominated by contributions from the regions around $K$ and $K'$. In particular, the scattering of right-handed into left-handed photons (left-handed into right-handed photons) mainly generates magnons at $K$ and $K'$ in the lower (upper) branch (see Fig.~\ref{fig:equilibrium}). This leads to a non-zero Raman circular dichroism that is shown below to be directly related to the Berry curvature. Since the Berry curvatures of the lower and upper branches are opposite, this also leads to a sign reversal of the RCD when the frequency crosses the gap.

Expanding the total Hamiltonian in powers of $\bA$ defines the $n$th-order light-matter couplings (LMCs) ${\bf L}^{(n)}$ as~\cite{Topp2021}
\begin{align}
 H(\bA) = H_0 + L_\mu^{(1)} A_\mu + L_{\mu\nu}^{(2)} A_\mu A_\nu + O({\bf A}^3).
\end{align}
The $n$th order LMC is a tensor of rank $n$ given by the $n$th order derivative of $H(\bA)$ with respect to $\bA$. When the dependence of the Hamiltonian on $\bA$ is of the form $H(\bk,\bA) = H(\bk - \bA)$, the derivatives with respect to $\bA$ can be replaced by derivatives with respect to $\bk$. In this case the linear and quadratic LMCs can be written as $L_\mu^{(1)} = -\partial_{k_\mu} H_0$ and $L_{\mu\nu}^{(2)} = \partial_{k_\mu} \partial_{k_\nu} H_0$. 

The relationship to the quantum geometric tensor is established by evaluating the matrix elements of the LMCs in the magnon basis. The magnon Hamiltonian for the canted honeycomb AFM has the general form $H = \sum_\bk \Psi_\bk^\dagger H_\bk \Psi_\bk$, where $\Psi^\dagger = (\alpha_\bk^\dagger, \beta_\bk^\dagger, \alpha_{-\bk}, \beta_{-\bk})$~\footnote{The discussion presented here has a straightforward generalization to any quadratic boson Hamiltonian $H = \sum_\bk \Psi_\bk^\dagger H_\bk \Psi_\bk$ with $L$ creation and $L$ annihilation operators written in the form of a spinor $\Psi^\dagger = (\alpha_{1\bk}^\dagger, \alpha_{2\bk}^\dagger, \ldots, \alpha_{L\bk}^\dagger, \alpha_{1,-\bk}, \alpha_{2,-\bk}, \ldots, \alpha_{L,-\bk})$.}. Since the geometry of the quantum states is encoded in the matrices $U_\bk$ connecting the magnon basis $\Psi$ to the spin-flip basis $\Phi$, it is useful to rewrite this as $H = \sum_\bk \Psi_\bk^\dagger ( \bar{U}_\bk \mathcal{H}_\bk U_\bk ) \Psi_\bk$, where $\mathcal{H}_\bk$ is the Hamiltonian in the $\Phi$ basis. The magnon Hamiltonian has a block structure, with matrix elements belonging to either of two categories: The first category corresponds to interband processes $\langle \alpha_\bk | H |\beta_{\bk} \rangle = (\bar{U}_\bk \mathcal{H}_\bk U_\bk)_{\alpha\beta}$, which are the only types of transitions allowed in the FM phase and will be denoted as FM processes. The second category corresponds to pair creation or annihilation processes $\langle \alpha_\bk \beta_{-\bk}| H |0\rangle = (\bar{U}_\bk \mathcal{H}_\bk U_\bk)_{\alpha\bar{\beta}}$, which are the only types of transitions allowed in the AFM phase and will be denoted as AFM processes. In the canted phase both FM and AFM processes contribute to the LMCs, and to distinguish them we use a bar over the index corresponding to a state with negative momentum. 

The LMCs are obtained from the Schr\"odinger equation $\mathcal{H}_\bk U_\bk = \mathcal{E}_\bk U_\bk$, where $\mathcal{E}_\bk$ is the diagonal matrix of eigenvalues of $\tau_z\mathcal{H}_\bk$ and $\tau_z$ is the third Pauli matrix in Bogoliubov space. Differentiating this equation and multiplying by $\bar{U}_\bk$ from the left gives $\bar{U}_\bk [\partial_\mu \mathcal{H}_\bk] U_\bk = [\partial_\mu \mathcal{E}_\bk] \bar{U}_\bk U_\bk + (\mathcal{E}_\bk - \bar{\mathcal{E}}_\bk) [\bar{U}_\bk \partial_\mu U_\bk]$, where $\mathcal{E}_\bk$ and $\bar{\mathcal{E}_\bk}$ are used to distinguish energies corresponding to columns (rows) of $U_\bk$ ($\bar{U}_\bk$). The quadratic LMCs are similarly obtained from the second derivatives of the Schr\"odinger equation, and the matrix elements of the linear and quadratic LMCs are summarized in Tab.~\ref{tab:quantum_geometry}. In particular, the quadratic LMCs are found to be related to the quantum geometric tensor $T_{\mu\nu}^n = \langle \partial_\mu n_\bk| (1 - |n_\bk\rangle\langle n_\bk|) |\partial_\nu n_\bk\rangle$~\cite{Berry1989} expressed in terms of the matrix $U_\bk$. Since the above argument only relies on the form of the Hamiltonian and the relation $H(\bk,\bA) = H(\bk - \bA)$, the expressions in Tab.~\ref{tab:quantum_geometry} hold for any quadratic bosonic Hamiltonian with this property.

Although the formal expressions are identical for FM and AFM processes, the application of $\mathcal{H}_\bk$ to the transformation matrices $U_\bk$ introduces a factor $\tau_z$ in the AFM case (since $u_{n\bk}$ are eigenvectors of $\tau_z\mathcal{H}_\bk$). This changes the meaning of $\bar{u}_{n\bk}$ from being a column of the Hermitian transpose $U_\bk^\dagger$ to a column of the inverse $U_\bk^{-1}$ (in the FM case these matrices are equivalent).

%%%%%%%%%%%%%%%%%%%%%%%%%%%%%%%%%%%%%%%%%%%%%%%%%%%%%%%
%%%%%%%%%%%%%%%%%%%%%%%%%%%%%%%%%%%%%%%%%%%%%%%%%%%%%%%
%%%%%%%%%%%%%%%%%%%%%%%%%%%%%%%%%%%%%%%%%%%%%%%%%%%%%%%

%\subsection*{Raman circular dichroism and Berry curvature}
The general relations in Tab.~\ref{tab:quantum_geometry} can be used to establish a connection between the quantum geometric tensor and the Raman circular dichroism (RCD)~\cite{Jin2020,Xie2021}. The RCD is defined at normal incidence as $\chi = \mathcal{P}^R - \mathcal{P}^L$. It measures the difference in total scattering cross-section between an incident laser with right- and left-handed polarization, denoted by $\mathcal{P}^R$ and $\mathcal{P}^L$, respectively. The scattering cross-section due to the Raman Hamiltonian $H_R$ is given by
\begin{align}
 \mathcal{P}^s = \sum_{ns'} |\langle \Psi_n|H_R^{ss'}|\Psi_0\rangle|^2 \delta(\hbar\omega + E_0 - E_n),
\end{align}
where $H_R \propto ({\bf e}_s^{*}\cdot{\bf d}_{ij})({\bf e}_{s'}\cdot{\bf d}_{ij})$, ${\bf e}_s$ is a photon polarization vectors, and ${\bf d}_{ij}$ is the vector between spins $i$ and $j$. The incident polarization is denoted by $s = R$ ($s = L)$ for right-handed (left-handed) light, and the scattered polarization $s'$ is summed over. Further, $|\Psi_n\rangle$ are eigenstates of the equilibrium Hamiltonian $H_0$, and the Raman energy $\hbar\omega = \hbar\omega_{\rm in} - \hbar\omega_{\rm sc}$ is the difference between incident and scattered photon energies. The RCD can be written in terms of the LMCs as $\chi = 8 \im[ (L_{xx}^{(2)} - L_{yy}^{(2)}) \bar{L}_{xy}^{(2)} ]$, where $\bar{L}_{\mu\nu}^{(2)}$ denotes the complex conjugate of $L_{\mu\nu}^{(2)}$. 

In the collinear AFM limit $B \to 0$, the Berry curvature and the RCD are directly related via $\Omega = -\chi/(2a^2 d_\bk^2)$. Here $a$ is the lattice constant, ${\bf d}_\bk = (h_0, h_x, h_y)$, and $h_i$ is the component of the Hamiltonian proportional to the Pauli matrix $\sigma_i$. However, for the collinear AFM both the Berry curvature and the RCD are independent of the DMI, and the system is topologically trivial. This follows from the fact that the AFM LMCs depend on the sum $\epsilon_{m\bk} + \epsilon_{n\bk}$ (cf. Tab~\ref{tab:quantum_geometry}), so that any dependence on $D$ cancels. The relation between $\chi$ and $\Omega$ further shows that frequency integrated RCD of a collinear AFM vanishes. In the FM limit $B > B_s$ the the Berry curvature and RCD are related by $\Omega = -\chi/(2a^2 d_\bk^2) + \rho_\bk/(2d_\bk^3)$, where ${\bf d}_\bk = (h_x, h_y, h_z)$ and $\rho_\bk$ is a term linear in $D$. For $D/J \lesssim 0.1$ this term is small, and the circular dichroism is approximately given by $\chi \approx - (2a^2) \int d\bk\, d_\bk^2 \Omega f(\epsilon_\bk)$, where $f(\epsilon_\bk)$ is the Bose-Einstein distribution. For the typical low-temperature scenario $f(\epsilon_\bk) = \delta(\bk)$ the circular dichroism vanishes, while at finite temperature a non-zero value might be assumed.

%%%%%%%%%%%%%%%%%%%%%%%%%%%%%%%%%%%%%%%%%%%%%%%%%%%%%%%
%%%%%%%%%%%%%%%%%%%%%%%%%%%%%%%%%%%%%%%%%%%%%%%%%%%%%%%
%%%%%%%%%%%%%%%%%%%%%%%%%%%%%%%%%%%%%%%%%%%%%%%%%%%%%%%

Fig.~\ref{fig:canted} shows the zero temperature RCD of the canted AFM as a function of DMI, external magnetic field and photon energy. Clearly, the integrated RCD is closely related to the Chern number of the magnon bands. It vanishes in the topologically trivial state but is non-zero otherwise, and thus constitutes an optical probe of the magnon band topology in canted AFMs. In this sense the Berry curvature determines the circular dichroism of a canted honeycomb AFM in a manner strongly reminiscent of the relationship between the circular dichroism and Berry curvature found in electronic systems~\cite{Yao08}.

Fig.~\ref{fig:frequency} further shows that performing frequency-resolved measurements of the RCD provides direct access to the topological gap $\Delta_g$ at $K$. The dominant Raman processes create magnon pairs at $K/K'$, and thus the RCD is dominated by the Berry curvature at these points. Since $\Omega_+ = -\Omega_-$ at $K$ and $K'$ the RCD changes sign as $\omega$ traverses the gap, and the distance $\Delta_{\rm RCD}$ between the negative and positive peaks of the RCD gives a direct measurement of the topological gap $\Delta_g$. As the sign of the Berry curvature is determined by the direction of the magnetic field $B$, the RCD changes sign when the magnetic field direction is inverted.

In addition to probing the topological gap, Fig.~\ref{fig:canted}d shows that the RCD gives a measure of the two-magnon density of states (2DOS) weighted by Berry curvature, defined as $\Lambda = \Omega_{\alpha} \rho_{\alpha\alpha} + \Omega_{\beta} \rho_{\beta\beta}$ where $\rho_{\mu\nu}(\epsilon) = \sum_\bk \delta(\epsilon - \epsilon_\mu - \epsilon_\nu)$. Assuming the 2DOS can be independently probed, the shape of the RCD (peak width and intensity) provides information about the $\bk$-space distribution of the magnon Berry curvature. In particular, it shows that for the canted AFM the main sources of Berry curvature are located at $K$ and $K'$, and that the Berry curvature changes sign between the magnon bands.

%%%%%%%%%%%%%%%%%%%%%%%%%%%%%%%%%%%%%%%%%%%%%%%%%%%%%%%
%%%%%%%%%%%%%%%%%%%%%%%%%%%%%%%%%%%%%%%%%%%%%%%%%%%%%%%
%%%%%%%%%%%%%%%%%%%%%%%%%%%%%%%%%%%%%%%%%%%%%%%%%%%%%%%

%\section{Discussion}\label{sec:discussion}
%\subsection*{Mechanisms of light-matter coupling}
We have shown that the magnon band topology of canted antiferromagnets is probed by the circular dichroism of the dominant two-magnon Raman scattering process. In particular, frequency-resolved RCD measurements give direct access to the topological gap as well as to the $\bk$-space distribution of the magnon Berry curvature, while the integrated signal is tied to the Chern number of the magnon bands. Since band topology in magnon systems is notoriously hard to validate, due to the bosonic exchange statistics of the quasiparticles and the lack of a chemical potential, these findings provide an important step towards utilizing topological magnon excitations in functional spintronics devices.

%%%%%%%%%%%%%%%%%%%%%%%%%%%%%%%%%%%%%%%%%%%%%%%%%%%%%%%
%%%%%%%%%%%%%%%%%%%%%%%%%%%%%%%%%%%%%%%%%%%%%%%%%%%%%%%
%%%%%%%%%%%%%%%%%%%%%%%%%%%%%%%%%%%%%%%%%%%%%%%%%%%%%%%

%\subsection*{Material realizations}
Our results are of relevance for a wide range of vdW magnetic insulators described by spin Hamiltonians such as Eq.~\ref{eq:spin_ham}. These systems present diverse magnetic orders including collinear out-of-plane ferromagnetism (monolayer CrI$_3$, CrCl$_3$ and VI$_3$), collinear out-of-plane and in-plane AFM order (MnPS$_3$ and MnPSe$_3$, respectively), and zigzag AFM order (FePS$_3$, FePSe$_3$ and NiPS$_3$). A canted AFM can thus be realized by applying an in-plane or out-of-plane magnetic field to MnPS$_3$ or MnPSe$_3$, respectively. Out of the above mentioned materials, CrI$_3$ in particular has been argued to display substantial out-of-plane DMIs of the appropriate form~\cite{Chen2018,Chen2021}.

%%%%%%%%%%%%%%%%%%%%%%%%%%%%%%%%%%%%%%%%%%%%%%%%%%%%%%%
%%%%%%%%%%%%%%%%%%%%%%%%%%%%%%%%%%%%%%%%%%%%%%%%%%%%%%%
%%%%%%%%%%%%%%%%%%%%%%%%%%%%%%%%%%%%%%%%%%%%%%%%%%%%%%%

%\subsection*{Discussion}
To establish the connection between the Raman circular dichroism and the magnon Berry curvature we have derived a general relation between the magnon LMCs and the quantum geometric tensor. Utilizing this relation we have shown that the magnon band topology of canted AFMs is probed by the RCD. However, since these relations hold for general quadratic boson Hamiltonians, our results pave the way for probing the quantum geometry of diverse bosonic quasiparticles such as magnons, photons, excitons and polaritons. In addition to probing magnon states, the two-magnon Raman processes studied can be used to generate magnons at the $K$ and $K'$ points. Since this is where topological edge modes are expected to appear in a finite geometry, it is likely that such processes can be used to generate magnon edge currents with tunable propagation~\cite{Bostrom2021}. Our work thereby opens vast possibilities for interfacing light and topological magnon modes in functional spintronics devices.

%%%%%%%%%%%%%%%%%%%%%%%%%%%%%%%%%%%%%%%%%%%%%%%%%%%%%%%
%%%%%%%%%%%%%%%%%%%%%%%%%%%%%%%%%%%%%%%%%%%%%%%%%%%%%%%
%%%%%%%%%%%%%%%%%%%%%%%%%%%%%%%%%%%%%%%%%%%%%%%%%%%%%%%

\begin{acknowledgments} 
%\section*{Acknowledgements} 
We acknowledge support by the Max Planck Institute New York City Center for Non-Equilibrium Quantum Phenomena, the Cluster of Excellence “Advanced Imaging of Matter” (AIM) and Grupos Consolidados (IT1249-19). S.~V.~K. acknowledges funding by the Max Planck Society in the form of a Max Planck Research Group and by the Deutsche Forschungsgemeinschaft (DFG, German Research Foundation) -- Project-ID 429529648 -- TRR 306 QuCoLiMa (``Quantum Cooperativity of Light and Matter''). M.~A.~S. acknowledges financial support through the Deutsche Forschungsgemeinschaft (DFG, German Research Foundation) via the Emmy Noether program (SE 2558/2). J.~W.~M acknowledges support from the Cluster of Excellence ‘CUI: Advanced Imaging of Matter’ of the Deutsche Forschungsgemeinschaft (DFG), EXC 2056, project ID 390715994, the Deutsche Forschungsgemeinschaft (DFG, German Research Foundation) – SFB-925 – project 170620586, and the Max Planck-New York City Center for Non-Equilibrium Quantum Phenomena. The Flatiron Institute is a Division of the Simons Foundation.

%\section*{Data Availability}
%All data supporting the findings of this study are available from the corresponding authors upon reasonable request.

%\section*{Code Availability}
%The codes used to generate the data of this study are available from the corresponding authors upon reasonable request.
\end{acknowledgments}

% \section*{Author Contributions}
% 
% \section*{Competing Interests}
% The authors declare no competing financial or non-financial interests.

%%%%%%%%%%%%%%%%%%%%%%%%%%%%%%%%%%%%%%%%%%%%%%%%%%%%%%%
%%%%%%%%%%%%%%%%%%%%%%%%%%%%%%%%%%%%%%%%%%%%%%%%%%%%%%%
%%%%%%%%%%%%%%%%%%%%%%%%%%%%%%%%%%%%%%%%%%%%%%%%%%%%%%%

\bibliography{references_circular}

\end{document}